\documentclass{article}
\usepackage[dvips]{graphicx}
\usepackage[T1]{fontenc}
\usepackage{setspace}
\usepackage{graphicx}
\usepackage{amsmath}
\usepackage{bm}
\usepackage{amsfonts}
\usepackage{amssymb}
\usepackage{epsfig}
\usepackage{bookman}
\usepackage{latexsym}
\begin{document}
\begin{center}
{\Large\bf MOND virial theorem applied\\[5PT]
to a galaxy cluster\\[5PT]}
\vspace{0.6cm}
J. C. Fabris\footnote{e-mail: fabris@pq.cnpq.br} and H.E.S. Velten\footnote{e-mail: velten@cce.ufes.br. Present address:
Fakult\"at f\"ur Physik, Universität Bielefeld, Postfach 100131,
33501, Bielefeld, Germany} \medskip \vspace{0.5cm}

{ \it Departamento de F\'{\i}sica, Universidade Federal do Esp\'{\i}rito Santo, Vit\'oria, Esp\'{\i}rito Santo, Brasil}\medskip

\abstract
{Large values for the mass-to-light ratio $(\Upsilon)$ in self-gravitating systems is one of the most important evidences of dark matter. We propose a expression for the mass-to-light ratio in spherical systems using MOND. Results for the COMA cluster reveal that a modification of the gravity, as proposed by MOND, can reduce significantly this value.}

{\bf Keywords:} MOND, COMA cluster, Dark Matter.
\end{center}
\section{Introduction}
The standard model for cosmology states that $95\%$ of all content of the universe belongs to a dark sector, not yet detected, composed by a unknown form of energy called dark energy and some kind of non-baryonic matter or dark matter, while baryogenesis implies that only $5\%$ of the universe is made up of atoms. In general terms, dark matter is the
unknown fraction that appears in agglomerated matter, like
galaxies and clusters of galaxies, having a effective null
pressure and suffering consequently the clustering process. Dark
energy is a smooth component of the matter content of the Universe
which must have negative pressure, inducing the accelerated
expansion of the Universe. A review of the observational evidences
and theoretical explanations for dark matter and dark energy can
be found in references \cite{han1,pad1,sahni1}.
\par
The first evidence on the existence of dark matter came from the
analysis of the dynamics of the Coma cluster of galaxies, in the
thirties \cite{zwicky}. Using the virial theorem and the
hypothesis that the Coma cluster is in dynamical equilibrium, it
was found that there should have much more matter than it could be
deduced using the usual mass-to-light ratio for galaxies.
Later, the study of the rotation curve of spiral galaxies has
shown that even at the level of galaxies the major part of matter
does not emit any kind of radiation. In galaxies, the
mass-to-light ratio using the rotation curves can be as high
as 30, in solar unities, while this number is of the order of 180 for the Coma
cluster \cite{kent}.
\par
In general, the existence of dark matter is proposed in order
to explain anomalies observed in the dynamics of self-gravitating systems,
like galaxies and clusters of galaxies. An alternative to dark matter is
to admit that the fundamental laws of physics are not the usual one from a certain scale on. This
can be achieved, for example, by changing the gravitational potential. Another possibility,
at newtonian level, is to modify the Newton's dynamic laws. One interesting proposal in
this sense is the MOND theory \cite{milgrom1}, which is based on a changing of the Newton's second law, such that
it takes the form
\begin{equation}
m\mu\biggr(\frac{a}{a_0}\biggl)\vec a = \vec F \quad ,
\end{equation}
where $\mu(x)$ is a function such that $\mu(x) \approx 1 $ for $x >> 1$ and $\mu(x) \approx x$ for $x << 1$.
Hence, $a_0$ defines a critical acceleration below which the usual Newton's second law is not
valid anymore. This phenomenological proposal has been very succesfull in explaining the rotation curve
of the spiral galaxies, $a_0$ taking a universal value around $a_0 \sim 10^{-8}\,cm.s^{-2}$ \cite{sanders}. Some interpolation
formula between the "newtonian" regime and the "MOND" regime for $\mu(x)$ have been proposed in the literature \cite{milgrom2}.
\par
One of the main criticism with respect to the MOND theory is that it is non-relativistic, and it can not
be deduced from a Lagrangian. More recently, a relativistic version of the theory has been proposed, involving,
beside the usual geometric and matter sector, a vector field and a scalar field \cite{bekenstein}. But, it remains some controversy if MOND can succesfully explain the lensing effect in clusters of galaxies \cite{lentes}.
\par
The proposal of the present work is to go back to the dynamics of cluster of galaxies, in the realm of the MOND theory.
There has been some studies in this sense using X-ray data \cite{moffat}. The purpose of the present work is more modest: to generalise the
virial theorem using a suitable interpolation expression for $\mu(x)$ and from it to deduce the mass-to-light ratio.
To be specific, we work with a given set of observational data concerning the Coma cluster.
It will be shown that, even if a closed expression for the generalised virial formula is not possible, we can obtain the
mass-to-light ratio that is compatible with ordinary estimations without dark matter: values as low 10 can be easily obtained.
The paper is organised as it follows. In next section, we present the general expressions for virial theorem. In section $3$ the observational
data for the Coma cluster are discussed and a estimation for the mass-to-light ratio is obtained. In the last section we present our
conclusions.

\section{Method}

The general form of the MOND theory is given by
\begin{equation}
\label{e1}
m\mu(x)\vec a = - G\frac{M\,m}{r^2}\hat r \quad ,
\end{equation}
where $\mu$ is a function, to be specified, which has the following properties: $\mu(x) \sim 1$, for $x >> 1$, and $\mu(x) \sim x$ for $x << 1$. Since $x = a/a_0$, where
$a_0$ is a constant with dimension of acceleration, the MOND theory implies the introduction of a new fundamental constant: a critical acceleration $a_0$ which specifies
when the newtonian mechanics is valid.
Expression (\ref{e1}) describes the motion of a point mass $m$ under the influence of the gravitational
field of a spherical large mass distribution with total mass $M$.
The function $\mu(x)$ is arbitrary, excepted by the fact it must obey the assymptotic regimes described before.
We will adopt the following form for this function:
\begin{equation}
\label{function}
\mu(x) = \left\{1+\left(2x\right)^{-2}\right\}^\frac{1}{2} -
\frac{1}{2x} = \frac{1}{2x} \left\{\sqrt{1+\left(2x\right)^2} -
1\right\}\quad ,
\end{equation}
with
\begin{equation}
\vec x = \frac{\vec a}{a_0} \quad .
\end{equation}

\begin{figure}
\begin{center}
\includegraphics[width=0.4\textwidth]{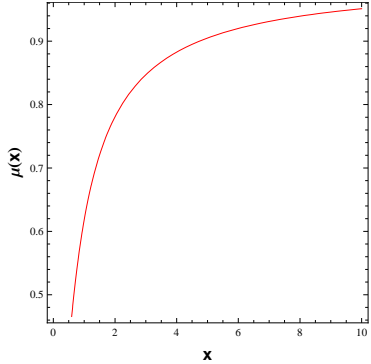}
\caption{Behaviour of the function $\mu(x)$. The newtonian limit is reached when $\mu \rightarrow 1$.}
\end{center}
\end{figure}

\par
For technical reasons, it is convenient to rewritte equation (\ref{e1}) as
\begin{equation}
\vec x = -G\frac{M}{\mu(x) r^2 a_0}\hat{r}.
\end{equation}
Some simple manipulations lead to the expression
\begin{equation}
\label{e2}
\vec a_{M} = -\frac{GM}{r^2}\sqrt{1+\frac{r^2 a_0}{GM}}\hat{r} \quad .
\end{equation}
With the identification $\sqrt{GM/a_0} = r_c$, where $r_c$ is a "critical distance" beyond which the
MOND regime becomes effective, equation (\ref{e2}) becomes
\begin{equation}
\vec a_{M} = -\frac{GM}{r^2}\sqrt{1+\frac{r^2}{r_c^2}}\hat{r} \quad .
\end{equation}
It must be stressed that the MOND critical parameter is the acceleration scale $a_0$.
However, with the formal definition of a distance scale, the MOND theory is recast as a (formal) modification of the gravitational potential.
The new form of the gravitational potential is dictated by the choice of the function $\mu(x)$. We will use
this approach from here on, but the results does not depend on it: it is just a question of description of
the dynamics. Using such approach the potential corresponding to the function $\mu(x)$ chosen is
\begin{equation}
\label{potential}\Phi_{M}\left(r\right) = -GM\left\{\frac{\sqrt{1+\left(\frac{r}{r_c}\right)^2}}{r} -
\frac{\sinh^{-1}(\frac{r}{r_c})}{r_c}\right\}.
\end{equation}
The potential has the usual newtonian form in the limit $r << r_c$, while it becomes logarithmic 
when $r >> r_c$, what corresponds to the full MOND regime.
\par
The first indirect detection of dark matter came from the application of the virial theorem, which is
valid in the Newtonian dynamics, to the Coma cluster \cite{zwicky}. This theorem establishes that in a self-gravitating system,
in dynamical equilibrium, the potential energy is minus two times the kinetic energy. Measuring the kinetic energy, what can
be done through a spectroscopic analysis of the light coming from a distant system like a cluster of galaxies, the potential energy, and
hence the mass of the system, can be deduced. But, such simple relation between the potential and kinetic energy can only be possible
in the case of a newtonian theory. For a different gravitational potential (or a different dynamic law), the virial theorem in its usual
form is not valid. The general relation for the virial theorem assumes the form 
\begin{equation}
\label{virial}
K = \frac{1}{2} \sum_{i} \left\langle \frac{\partial \Phi(r)}{\partial
r}r\right\rangle{m_{i}} \quad,
\end{equation}
where it was computed the contribution from all {\it i}-components of the system with mass $m_{i}$. Introducing (\ref{potential}) in the above relation the kinetic energy of a self-graviting system orbiting in the full MOND regime can be obtained as
\begin{equation}
\label{kinetic}
K=\frac{G M(r)}{2r}\sqrt{1+\frac{r^{2}}{r^{2}_{c}}}\,\,m,
\end{equation}
where $m$ is the mass of a tiny shell at radius $r$, and $M(r)$ is the mass enclosed by this shell.
\par
If we know the kinetic energy of a spherical system it is possible to evaluate the mass-to-light ratio $\Upsilon$ under the assumption that $\Upsilon$ is independent of radius. The kinetic energy can be associated with the surface luminosity $I$ and the line-of-sight velocity dispersion $\sigma$ by
\begin{equation}
\label{masstolight} K=\Upsilon J\,\,,
\end{equation} 
where $J$ is a integral defined as \cite{binney}
\begin{equation}
\label{J}J=3\pi \int_{0}^{\infty}I(R)\sigma^{2}(R) R dR.
\end{equation}    
Moreover, a symmetric mass distribution should be described by some kind of density profile. The mean density may be associated with the observational data and can be written as \cite{binney}
\begin{equation}
\label{density}\rho\left(r\right)=-\frac{\Upsilon}{\pi}\int^{\infty}_{r}\frac{dI}{dR}\frac{dR}{\sqrt{R^{2}-r^{2}}}.
\end{equation}
The above definition enable us to calculate the expression for the mass-to-light ratio with MOND. Inserting (\ref{density}) twice in the MOND expression for the kinetic energy (\ref{kinetic}) and combining this result with (\ref{masstolight}) it is possible to determine the mass-to-light ratio using only the velocity dispersion and the surface luminosity data:
\begin{equation}
\Upsilon_{M}=-\frac{2J}{\widetilde{J}}.
\end{equation}
In above relation $J$ is the integral define before and $\widetilde{J}$ is defined as
\begin{equation}
\label{Jtil}\widetilde{J}=-16G\int^{\infty}_{0} t(r)p(r)r\sqrt{1+\frac{r^{2}}{r^{2}_{c}}} \,\,dr,
\end{equation}
where $t(r)$ and $p(r)$ are also auxiliar integral expressions that depend on $I$ and $\sigma$
\begin{equation}
t(r)=\int^{\infty}_{0}\frac{dI}{dR}\frac{dR}{\sqrt{R^{2}-r^{2}}}\hspace{1cm}p(r)=\int^{r}_{0}t(r^{\prime}){r^{\prime}}^{2}dr^{\prime}.
\end{equation}
In order to calculate $\Upsilon$ for a self-gravitating system we need the surface luminosity profile $I(R)$ and the dispersion velocity profile $\sigma(R)$. Note that the velocity dispersion appears only in term $J$. Hence, $J$ represents the kinematic contribution from the virial theorem. We aply this result for the COMA cluster data in next section.
\par
\section{COMA Cluster Data Analysis}

The mass-to-light ratio of the COMA cluster was estimated, using the Newtonian theory, as $\Upsilon_{N}\approx 180$ \cite{kent}. It seems reasonable to interpret this large value for the mass-to-light ratio as a strong evidence in favour of dark matter. If MOND theory provide a succesfully description for the dynamics of this cluster we hope to find for $\Upsilon$ a value at least one order of magnitude less than $180$. However, before any tentative to reconstruct a analysis in the COMA cluster using MOND theory, we need to assume some properties about the cluster:
\begin{enumerate}
\item We adopt a spherical symmetry. Although COMA have a elongated form, we are interested in the radial structure of the cluster.
\item Any dark component of the system traces the luminous matter. With this assumption we consider a single component model for the cluster. This condition can be relaxed, however we will not consider this case. 
\item The cluster is in a stage of dynamical equilibrium and all the mechanics of the cluster is well known and not yet relativistic.
\end{enumerate}
\par 
The above assumptions don't reflect actual gravitational systems, however will allow us to obtain in a simple way the cluster mass-to-light ratio. For a MOND description of orbits in non trivial symmetries see \cite{bullet}.
\par
The COMA cluster has been so widely studied over the years and for this reason there are a lot of available observational data in the literature. In order to obtain the velocity dispersion and the surface luminosity profile we take all the cluster members and binning them radially. We use the
velocity dispersion data described in reference \cite{kent}. Each bin contains aproximately 20 members of the cluster. The equivalent data for each bin and the best fit for the velocity dispersion and the surface luminosity are showed in figure 1. We do not adopt, \emph{a priori}, any dynamical model for fitting the data. The best fit curves showed in figure 2 correspond to the curve that minimize the quatity $\chi^{2}$ that is proportional to the difference between the fitting value and the observational data in each bin.
Inserting the surface luminosity profile obtained from the best fit, as we can see in figure 1, in the auxiliar expressions (16) we can compute the integral $\tilde{J}$, Eq. (\ref{Jtil}). With the velocity dispersion profile we can evaluate the kinetic term $J$ in integral (\ref{J}). The mass-to-light ratio depends on the value of the critical radius $r_{c}$. Table 1 shows results for $\Upsilon_{M}$ for various values of $r_{c}$.

\begin{center}
\begin{figure}[!t]
\begin{minipage}[t]{0.38\linewidth}
\includegraphics[width=\linewidth]{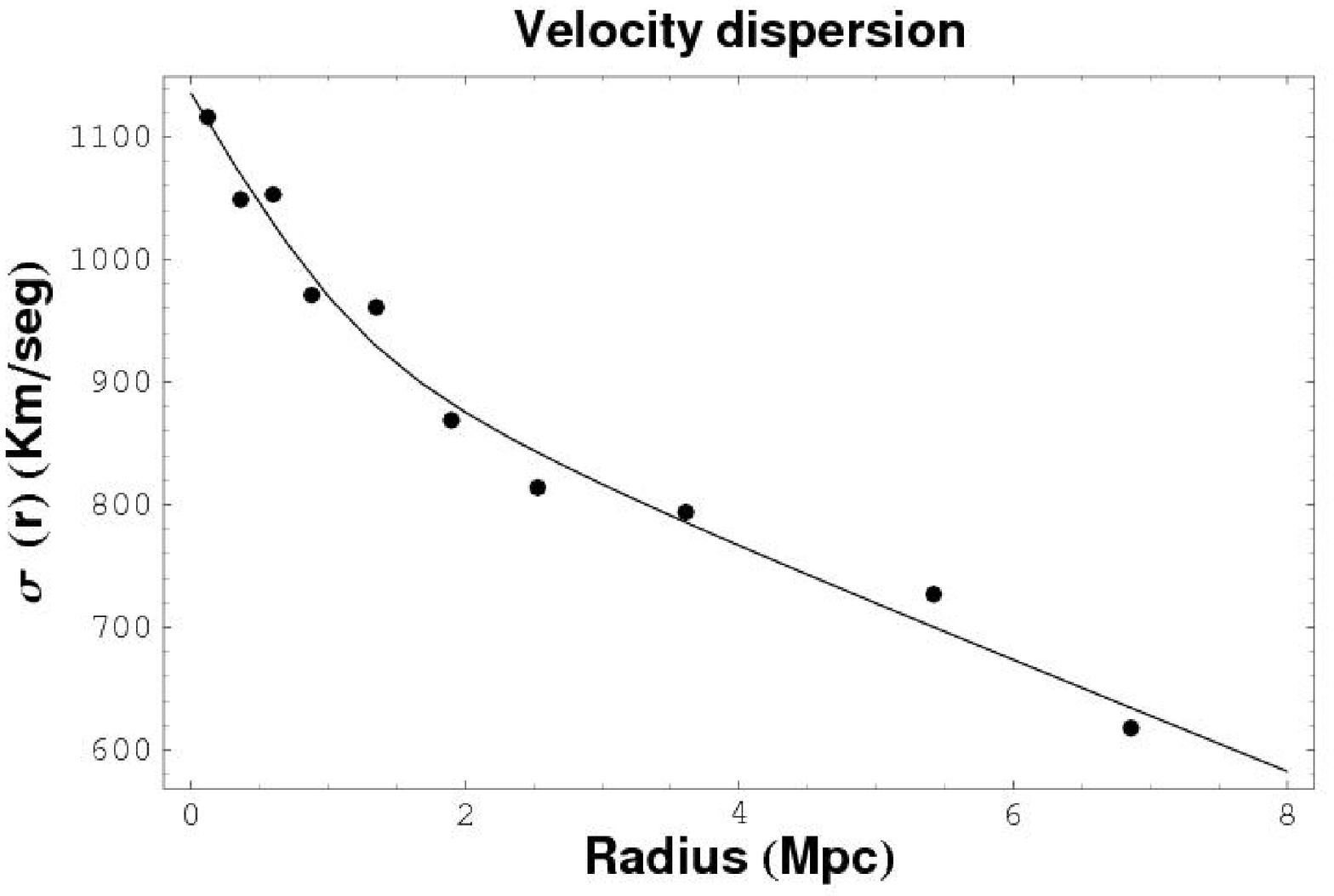}
\end{minipage} \hfill
\begin{minipage}[t]{0.4\linewidth}
\includegraphics[width=\linewidth]{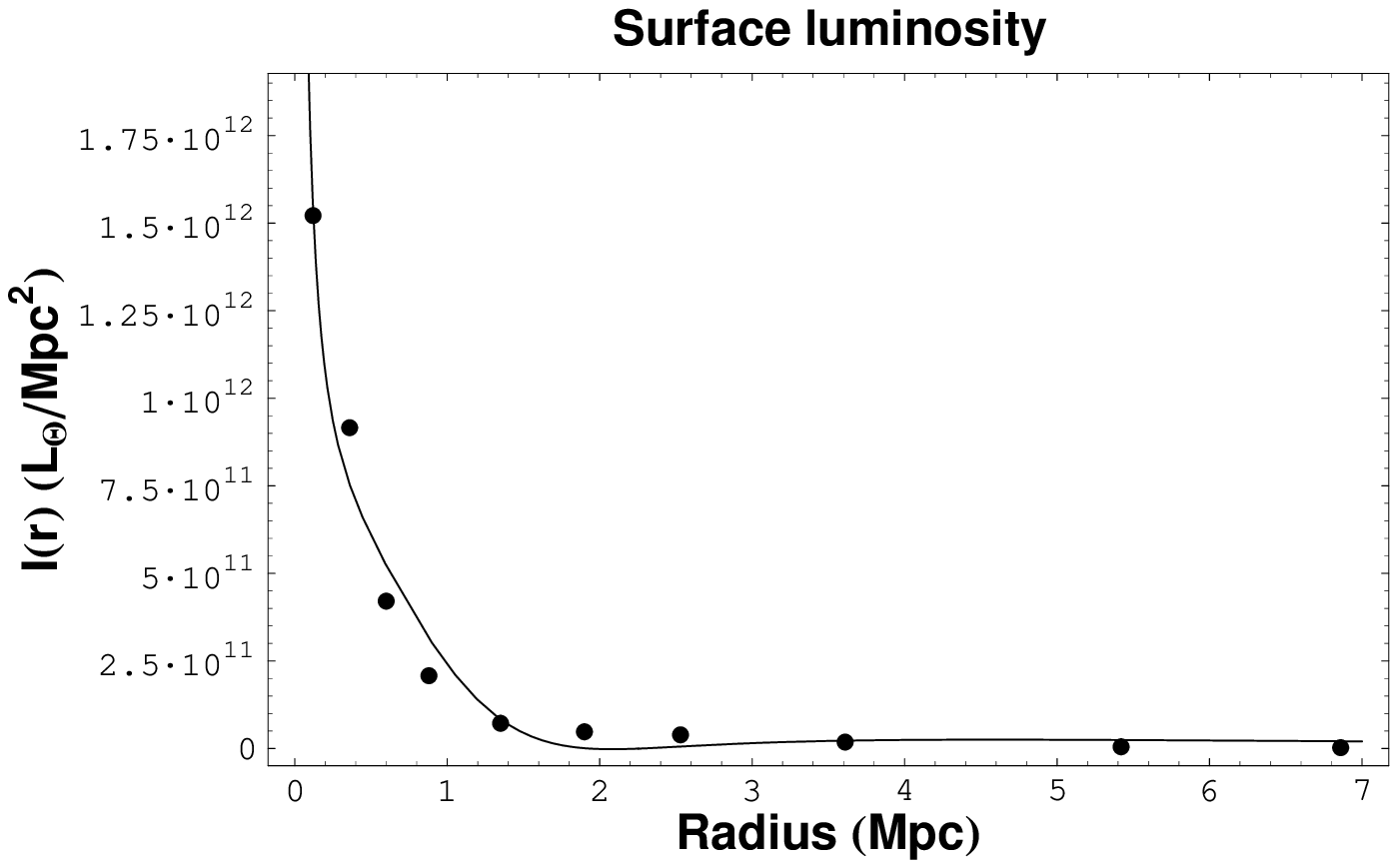}
\end{minipage} \hfill
\caption{Best fit for the velocity dispersion and surface luminosity profiles.}
\end{figure}
\end{center}

\begin{table}[b!]
\begin{center}
\begin{tabular}{l|c|r}
$r_{c} \left(Mpc\right)$   & $\Upsilon_{M}$ &  $a_{0}\left(m/s^{2}\right)$\\
\hline
     0.05 & 3.6  & 1.7$\times 10^{-7}$\\
     0.1  & 7.2  & 7.0$\times 10^{-7}$\\
     0.2  & 14.4 & 2.7$\times 10^{-8}$\\
     0.3  & 21.6 & 1.5$\times 10^{-8}$\\
     0.4  & 28.6 & 9.7$\times 10^{-9}$\\
     0.5  & 35.5 & 6.8$\times 10^{-9}$\\
     0.6  & 42.3 & 5.1$\times 10^{-9}$\\
     0.7  & 48.9 & 4.0$\times 10^{-9}$\\
     0.8  & 55.3 & 3.1$\times 10^{-9}$\\
     0.9  & 61.6 & 2.6$\times 10^{-9}$\\
\end{tabular}
\end{center}
\caption{Critical acceleration parameter $\left( a_{0}\right)$ and mass-to-light ratio in MOND theory $(\Upsilon_{M})$. We include the corresponding
values for the critical radii $r_{c}$, an auxiliar quantity derived from $a_0$. The Newtonian analysis $(r_{c}\rightarrow\infty)$ reveals for COMA cluster a value for the mass-to-light ratio $\Upsilon_{N}=208.4$}
\end{table}  
\section{Final Remarks}
The nature and origin of the dark matter is still one of the greatest challenges in cosmology and astrophysics. Its influence in the dynamics of galaxies and clusters and its distribution in these systems has been studied over the years and we don't have yet a satisfactory description for this unknown component. In this contribution we apply the MOND theory in order to obtain a relation for the mass-to-light ratio in self-gravitating systems. Only with surface luminosity and dispersion velocity data from COMA cluster the value of mass-to-light ratio was reduced using the MOND theory when compared with the Newtonian one. The exact value of $\Upsilon_{M}$ depends on the critical radii $r_{c}$. Table 1 shows several results of our analysis. For suitable values of $r_{c}$ we found lowers mass-to-light ratios in contrast with the previous Newtonian result $\Upsilon_{N}\approx 180$\cite{kent}. The Newtonian limit of our relations $(r_{c}\rightarrow\infty)$ indicates $\Upsilon_{N}=208.4$ for the COMA cluster. We can admit that the estimation of the mass-to-light ratio as considered above is highly uncertain, however MOND correct the huge Newtonian mass-to-light ratio down to values compatible with baryonic matter alone. In particular, using the critical acceleration $a_0 \sim 10^{-10}\, m/s^2$, a value that
leads to a good fit of the rotation curve of spiral galaxies in the MOND theory \cite{sanders}, a mass-to-light ratio of order of $70$ is obtained, a third
of the Newtonian value. In the analysis presented here, we have not taken into account the gas in the intergalactic medium of the
cluster, which is typically responsible for $20\%$ of the total mass of a cluster \cite{ettori}.
Even so, it agrees with the evaluation made in reference \cite{sandersbis} where X-ray data are taken into account. Note, however, that our results disagree with the simplified estimations
made in \cite{milgrom1} where just some general properties of the MOND theory have been used, leading to a very low 
mass/luminosity ratio, typically of order of a few tenths. Even if the necessity of dark matter remains, the dark matter problem is
alleviate. Moreover, since there are many simplifications in the analysis done (the hypothesis of a continuous matter distribution,
the choice of a spherical symmetry for the cluster, etc.), the results obtained using the interpolation function (\ref{function}) in the context of the MOND
theory justify the investigation of a more realistic scenario, perhaps including a discrete distribution of the matter besides a continuous
gas distribution. Of course, such more realistic analysis would imply the necessity of numerical simulations.

\noindent
{\bf Acknowledgment:} We thank CNPq (Brazil) and FAPES (Brazil) for partial financial support. The remarks made by the anonymous referee
has permitted us to improve the text and the interpretation of the results.


\begin{thebibliography}{99}

\bibitem{han1} S. Hannestad, {\it Dark energy and dark matter
from cosmological observations}, astro-ph/0509320.

\bibitem{pad1} T. Padmanabhan, {\it The dark side of the
Universe}, in Proceedings of the 29th International Cosmic Ray Conference {\bf 10},  
47(2005).

\bibitem{sahni1} V. Sahni, Lect. Notes Phys. {\bf 653}, 141(2004).

\bibitem{zwicky} F. Zwicky, Helv. Phys. Acta {\bf 6}, 110(1933).

\bibitem{kent} S. M. Kent and J. E. Gun, Astronomical Journal {\bf 87}, 7(1982).

\bibitem{milgrom1} M. Milgrom, Astrophys. J. {\bf 270}, 365(1983); Astrophys. J. {\bf 270}, 371(1983); Astrophys. J. {\bf 270}, 384(1983).

\bibitem{sanders} R.H. Sanders and S. C. McGaugh, Ann. Rev. Astron. Astrophys. {\bf 40}, 263(2002).

\bibitem{milgrom2} M. Milgrom, {\it The modified dynamics: a status review}, in {\bf Dark matter in astrophysics and particle physics, 1998},  
proceedings of the Second International Conference on Dark Matter in  
Astrophysics and Particle Physics, Heidelberg, Germany, 20-25 July,  
1998, edited by H.V. Klapdor-Kleingrothaus and L. Baudis, Institute of Physics Pub., Philadelphia(1999).

\bibitem{bekenstein} J.D. Bekenstein, Phys. Rev. {\bf D70}, 083509(2004); Erratum-ibid. {\bf D71}, 069901(2005).

\bibitem{lentes} H.S. Zhao, D.J. Bacon, A.N. Taylor and K. Horne, Month. Not. R. Astron. Soc. {\bf 368}, 171(2006); M-C. Chiu, C-M. Ko and Y. Tian, Astrophys. J. {\bf 636}, 565(2006); I. Ferreras, M. Sakellariadou, M.F. Yusaf, Phys.Rev.Lett. {\bf 100} 031302 (2008). 

\bibitem{moffat} J. R. Brownstein and J. W. Moffat, Month. Not. R. Astron. Soc. {\bf367}, 527-540 (2006); G.W. Angus, B. Famaey and D.A. Buote,
Month. Not. R. Astron. Soc. {\bf 387}, 1470(2008).

\bibitem{binney}
J. Binney and S. Tremaine, {\bf Galactic Dynamics}, Princeton University Press, Princeton (2004);  O. Tiret and F. Combes,
in {\bf  Formation and Evolution of Galaxy Disks},  
ASP Conference Series, {\bf 396}, Proceedings of the conference held  
1-5 October, 2007 at the Centro Convegni Matteo Ricci, Rome, Italy, edited by José G. Funes, S.J., and Enrico Maria Corsini, Astronomical Society of the Pacific, San Francisco(2008);  C. Nipoti, P. Londrillo e L. Ciotti, arXiv:0705.4633 (2007).

\bibitem{bullet}
G.W. Angus, B. Famaey, H.S. Zhao, Mon.Not.Roy.Astron.Soc. {\bf371}, 138 (2006).

\bibitem{ettori}  S. Ettori, S. De Grandi and S. Molendi, A\&A {\bf 391}, 841 (2002).

\bibitem{sandersbis} R.H. Sanders, Month. Not. R. Astron. Soc. {\bf 342}, 901 (2003).

\end{thebibliography}
\end{document}